\documentclass[reprint,amsmath,amssymb,aps,physrev,nofootinbib]{revtex4-2}
\usepackage{graphicx}
\usepackage{dcolumn}
\usepackage{bm}
\usepackage{url}
\expandafter\def\expandafter\url\expandafter{\url\Urlbreak}

\usepackage{multirow}
\usepackage{tikz}
\usepackage{stmaryrd}
\usepackage{subcaption}
\usepackage{orcidlink}
\usepackage{floatrow}
\usepackage{ulem}
\usepackage{soul}
\usepackage[french,english]{babel}
\usepackage[title]{appendix}
\usepackage{array}
\usepackage{tabularx}
\usepackage{hyperref}
\usepackage[noabbrev,capitalise]{cleveref}
\begin{document}
\preprint{APS/123-QED}

\title{\textbf{Connecting Electrical Grid Length and Material Stock to Population Density: Comparison of a French-Calibrated Scaling with 35 Electrified Countries}}

\author{Emile Emery\orcidlink{0000-0001-5074-1163}}

\author{Joseph Le Bihan \orcidlink{0009-0003-0736-4632}}

\author{José Halloy \orcidlink{0000-0003-1555-2484}}

\affiliation{Université Paris Cité, CNRS, LIED UMR 8236, Paris, F-75006, France}

\date{\today}

\begin{abstract}
The expansion of global electricity distribution systems necessitates the deployment of massive infrastructure. Assessing its implications from a spatial and material perspective requires an understanding of the core drivers of a distribution grid configuration. Our model samples substation locations using a non-linear relationship with population density and constructs a proxy network applying the Kruskal algorithm. This streamlined approach generates a proxy grid layout at the local scale and provides reasonable aggregate estimates of the total network length at the national scale. Using highly granular population data, this local model reveals a connection between population spread and distribution grid, which appears to persist at larger scales. Potentially driven by the emergent properties of population scaling laws, the aggregated network characteristics appear to be well described by multivariate power laws on aggregated population and area. Benchmarked against reported aggregate data for 35 countries, these results provide new multi-scale tools for characterizing electrical infrastructure and reveal key determinants of distribution grid extent. Combining network length estimates with material intensity data, we derive country-scale inventories of copper invested in medium-voltage lines, bounded by aerial and underground cabling assumptions. The same scaling law directly yields a macro-level proxy for copper mass as a function of population and area, extending the framework toward prospective assessments of material demand in electricity grid expansion.
\end{abstract}
\maketitle

\section{Introduction}

Electricity transmission and distribution networks contain a substantial share of the in-use stocks of aluminum and copper within society \cite{kalt_material_2021}. This share is expected to increase by an order of magnitude over the coming decades as electrification grows and electricity consumption rises \cite{IEA}.
Characterizing these networks has therefore become a critical issue at both local and global scales. Understanding their local spatial configuration—\textit{i.e.} the placement of substations and the branching of cables—is necessary to estimate both deployment costs and maintenance requirements. At the national scale, these local structures determine the total network length, which is the primary determinant of the grid's material content and economic cost. Length estimates are therefore essential for anticipating the resource investments associated with future grid expansion, as well as for assessing existing in-use material stocks from a circular economy perspective. In this work, we investigate this local-to-global connection through the lens of population density, which we show to be a powerful and parsimonious driver of distribution grid geometry at both scales.\\

Electrical power systems are structured in hierarchical layers, each serving distinct contributions to the transport of electricity from production sites to end consumers \cite{crastan_reseaux_nodate}. The precise voltage delimitation for each layer may vary from country to country, but we provide here a quick overview to sum up the main principles. The transmission network operates at the highest voltages to transport power over long distances with minimal losses. It comprises two sublayers: Extra High Voltage (EHV) networks (typically 200-1000 kV) that interconnect major production centers with major consumption areas such as administrative regions, and High Voltage (HV) networks (typically 45-200 kV) that transfer power from EHV to cities, industrial clusters, and large rural areas. The distribution network completes the delivery chain, operating at lower voltages to supply electricity to residential and small industrial consumers. It is divided into two distinct layers: the Medium Voltage (MV) network, operating at tens of kV, connects HV substations from the transmission layer to numerous local MV/LV substations and some industrial consumption nodes; and the Low Voltage (LV) network, operating around 400V, delivers power from these local substations directly to end users: households, factories, and commercial buildings.\\

The transmission segment of electricity networks is already relatively well documented, first because it involves much shorter total cable lengths, and second because it is largely determined by the spatial distribution of electricity generation sites \cite{emery_power_2025}. Moreover, this segment is generally managed by a limited number of large operators (Transmission System Operators, TSOs), which facilitates data aggregation and transparency. In contrast, the distribution segment covers significantly greater lengths and is often operated by a large number of local entities (Distribution System Operators, DSOs), making the collection and harmonization of local data considerably more complex. In addition, information on the location of power cables and substations can be sensitive, which further limits access to this data. France constitutes a notable exception, as a single DSO manages the vast majority of the distribution network. This distinction between transmission and distribution networks is also reflected in material composition \cite{surfer}, for physical and technical reasons.

Several methods have been employed to describe and model electricity distribution networks. This task is far from straightforward; Zvoleff \textit{et al.} notably demonstrated that an identical number of nodes, or substations, to be supplied can result in fundamentally different network configurations depending on the spatial distribution of their locations. The network could be composed mainly of short segments (nucleated population) or could require long cables between dispersed areas \cite{zvoleff_impact_2009}. Relying on satellite imagery, the authors demonstrated that rooftop detection can reliably estimate local low-voltage network lengths from rooftop detection, but this approach cannot distinguish electrified structures from non-electrified ones, and roofs are not equal in terms of electricity consumption. Similarly, Arderne et al. used night-time light intensity to predict the presence of medium-voltage lines at the regional scale, achieving 75\% accuracy at 15 km resolution, yet this method systematically underperforms in rural areas where light emissions remain below the detection threshold \cite{arderne_predictive_2020}.

More recently, full generative pipelines have demonstrated the 
large-scale production of geo-referenced, power-flow-compliant distribution grids from open data: Oneto \textit{et al.} generate the complete set of Swiss MV and LV grids with operational topologies validated against AC power-flow constraints \cite{oneto_large-scale_2025}, and Zapparoli \textit{et al.} build on these grids to allocate distributed energy resources with hourly resolution \cite{zapparoli_future_2025}. Such engineering-grade models achieve high electrical realism, at the cost of assembling multiple country-specific input layers (demand maps, transport and building data, equipment catalogs) and per-grid power-flow computations. The present work is of a different and complementary nature: it is not an engineering model but a descriptive, spatially explicit one, which infers aggregate geometric quantities---substation counts and network lengths---and their cross-country scaling from a single, globally available input, gridded population density, with a view to material-demand assessment.

The literature suggests that electricity networks tend to follow a set of general organizing principles, in particular the optimization of an overall system cost \cite{persoz_planification_1984}. This cost encompasses, among others, installation costs, Joule losses, reliability considerations and explains the observed topologies \cite{emery_power_2025,pagani_complex_2015}. While satellite imagery provides a direct way to identify the number and spatial distribution of grid-connected structures, it requires substantial processing effort. Population density appears to be a much simpler proxy for the spatial distribution of structures to be connected, and potentially a powerful one \cite{arderne_predictive_2020, dunn_spatial_2016}. This approach is particularly relevant for countries with a “mature” electricity network, where nearly $100\%$ of the population has access to household electricity. Although the theoretical derivation of the relationship between population density, the number of substations, and the various optimization heuristics governing network design remains challenging, the availability of detailed data for France makes it possible to propose a network model driven by population density as an input, combined with a parsimonious set of assumptions calibrated to reproduce the French distribution network. Population density captures where people reside rather than where electricity is effectively consumed. Two sources of discrepancy are worth noting: areas with high commercial or touristic activity can concentrate a significant number of grid connections despite a low permanent resident count; and light industrial facilities (whose heavy counterparts connect directly to the high-voltage layer) contribute to local network density independently of household distribution. Both effects are assumed negligible compared to the dominant contribution of the large-scale distribution of permanent households at the spatial scales considered here.

This relationship between population density and distribution network configuration seems to remain valid for other countries with full electrification. Multiplying total length by per-kilometer material intensities (with uncertainty bracketed by aerial vs. underground cabling assumptions) enables a first-order spatial allocation of in-use copper stocks at the national scale and reasonable aggregate estimates of future grid material demand. The strength of the framework is that it requires only population density as a spatial input, with a small set of calibrated geometric and material parameters. Population density data and projections are effectively far more accessible and well-developed than data on electricity distribution networks, which in some cases are simply unavailable. The central research question that we want to tackle can be stated as follows: to what extent is population density sufficient to reproduce the spatial structure and aggregate length of distribution networks ?\\

\subsection*{Objectives}

We aim to develop a spatially explicit generative model for the MV network, as well as simple scaling laws able to estimate the order of magnitude of the total cable length required for a complete electrical distribution network. The low-voltage (LV) layer is addressed only at the macro-scaling level. The main assumption is that electrical demand correlates well with population density in space. The existence of a link between population and network extent is well established \cite{zvoleff_impact_2009, arderne_predictive_2020, kalt_global_2021,
bettencourt_growth_2007, kuhnert_scaling_2006}.
This work aims to extend this understanding through the following objectives:

\begin{itemize}
    \item To develop a framework capable of generating a spatially explicit representation of the French MV electricity distribution network, using population density over the entire territory as the sole input, combined with a parsimonious set of parameters calibrated on French open data.

    \item To assess the accuracy of this model in the French case across several dimensions, including network topology, the spatial location of network nodes and edges, as well as aggregated metrics such as total medium-voltage line length or total number of nodes (substations).

    \item To evaluate and discuss the relevance of applying this model, calibrated on the French case, to other national MV grids.

    \item Based on the results obtained for other countries using this model calibrated on France, to discuss the invariants and the driving factors of aggregated quantities such as total medium-voltage and low-voltage line lengths, or the number of substations, in particular in relation to aggregated indicators such as population,
    land area, and electricity consumption.

    \item To translate network length estimates into country-scale inventories of copper invested in medium-voltage lines, using material intensity data, with a view to prospective assessments of material demand in grid expansion.
\end{itemize}

\section{Model}

\label{sec:Model}

Electricity distribution networks typically exhibit tree-like structures \cite{pagani_complex_2015}. Using this observation as a modeling assumption, a set of spatially explicit nodes (substations) can be transformed into a network. The key question then becomes how to model the number and spatial distribution of these nodes based on population density. We hypothesize the existence of a direct relationship between node density $n_{node}$ and population density $\rho$ across space : $n_{node}=\frac{N_{node}}{S_{cell}}=f(\frac{N_{pop}}{S_{cell}})=f(\rho)$. In the French case, comparing node density with population density at a fine spatial resolution ($S_{cell}$) makes it possible to identify and calibrate this relationship.\\

The WorldPop database provides an estimated population density per grid cell (people per $\text{km}^{2}$) for all countries, with a resolution of 30 arc seconds \cite{worldpop}. This data was produced using random forest-based machine learning techniques to disaggregate governmental population counts from administrative units into grid cells \cite{gaughan_spatiotemporal_2016,sorichetta_high-resolution_2015,stevens_disaggregating_2015}. 
The lattice has a fine resolution and captures the nucleation of households, which are crucial for network length measurements \cite{zvoleff_impact_2009}. These data will serve to calibrate the model on the French case and as input to provide a spatial representation of developed electrical distribution networks,  thereby estimating their total length for the other countries.\\

Regarding the set of nodes, the French DSO ENEDIS covers over 95\% of the French territory \cite{brochure_enedis} and provides an open database containing the locations of MV/LV substations from all DSO \cite{enedis_opendata}. Unfortunately, these nodes correspond only to substations connecting the MV network to the LV network, and the dataset does not include the terminal nodes or end-use structures, such as households, where the LV network ends. ENEDIS provides, in addition to data on network nodes, a set of line drawings (capturing the location of ENEDIS cables); however, these two datasets are not harmonized. Thus, the spatial accuracy of the line trajectories is insufficient to directly reconstruct the full network topology, \textit{i.e.}, the connectivity of substations \cite{venegas_electric_2021, emery_statistical_2025}. This lack of internal consistency further underscores the need for modeling approaches such as the one developed in the present study. We restrict our analysis, in a first step, to modeling the MV network, for which a complete set of nodes is available. Nevertheless, using population density to estimate the number and spatial distribution of terminal nodes appears to be a promising, potentially more effective approach.\\

The main difficulty in estimating node density as a function of population density is that WorldPop data are provided in a gridded spatial representation (square cells), which introduces potential edge effects. This induces a stochastic component in the relationship. In addition, the actual locations of substations themselves are likely to exhibit an intrinsic degree of randomness, as factors other than population density may locally influence their placement while having a negligible effect on average (for instance, the presence of a lake within an urban area preventing the installation of substations in part of a grid cell). Both sources of randomness are homogeneous and will be captured jointly using the Poisson law.\\

\begin{figure}[t]
    \centering
    \includegraphics[width=\linewidth]{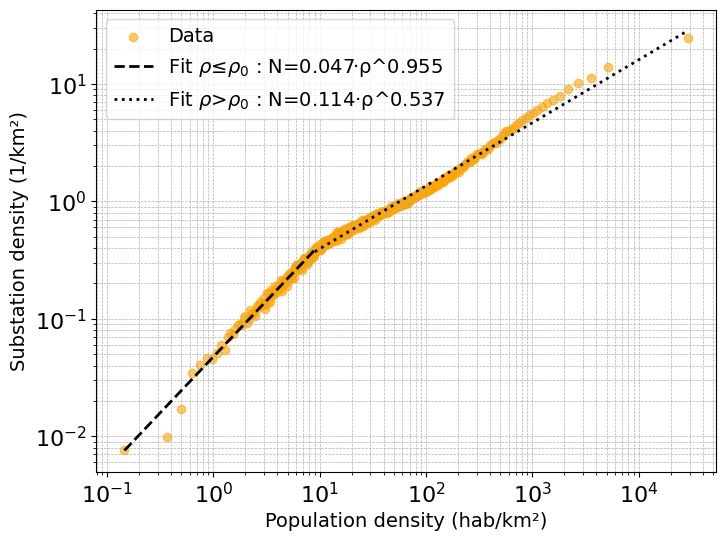}
    \caption{Empirical relationship between the number of substation in a cell and the corresponding population density. Each point corresponds to the average number of substations per bin of density. We built 500 bins containing the same number of data points. The threshold value is $\rho_0 \simeq 8.951$.}
    \label{fig:Nrho}
\end{figure}

Accordingly, \cref{fig:Nrho} depicts the average density of substations as a function of average population density, defined as the mean number of substations computed across all WorldPop cells corresponding to a given population density. This relationship, therefore, describes the expected value of the number of substations. Two distinct regimes are observed, with a break in slope around a critical density of approximately 9 inhabitants per $\text{km}^{2}$. The fitted relation is the following :

\begin{equation}\label{regimes}
    f(\rho) = 
    \begin{cases}
        \;0.05 \rho\text{ if }\rho < 9 \,\text{hab}/\text{km}^{2}\\
        \;0.1\rho^{0.5}\text{ if }\rho \geq 9 \,\text{hab}/\text{km}^{2}.
    \end{cases}
\end{equation}

This result is empirical but consistent with intuition: in sparsely populated areas, the number of substations is approximately proportional to the number of inhabitants, whereas in densely populated areas, scale effects emerge, making the number of substations markedly sublinear. This threshold value corresponds approximately to the transition threshold between an area considered \textit{uninhabited} and an \textit{inhabited} one in \cite{hanberry_imposing_2022}. This effect is closely related to the use of fixed-area density measures; reversing the perspective—by allocating a serviced area to each substation—eliminates this break (\textit{cf.} appendix \ref{vornoi_appendix}).\\

To account for the stochastic effects described above, and in particular to adequately represent the large number of cells containing no substations, we adopt a Poisson distribution. This choice is motivated by its simplicity, its spatial homogeneity, the absence of additional parameters, and its widespread use in counting processes. The data dispersion strengthens this validation with a variance-to-mean ratio of substation close to unity (value for a Poisson distribution) over most of the density range.
\cref{fig:prop_null}, which shows the proportion of cells without substations as a function of population density, supports the suitability of this probabilistic assumption.\\

The algorithm samples substations' locations cell by cell. For each cell $i$, a random draw $N_{i}$ is performed according to a Poisson distribution with parameter $n_{stat,i}=f(\rho_{i})$, where $\rho_{i}$ denotes the population density of the cell $i$ as provided by WorldPop, and $f$ is defined in \cref{fig:Nrho}. This draw determines the number of substations $N_{i}$ to be placed in cell $i$; these substations are then randomly and uniformly positioned within the cell. This Poisson distribution naturally models the discrete, random nature of node placement and the observed cell-to-cell fluctuations around the mean density, allowing for the recovery of a plausible node sprinkling from a grid-cell dataset (\textit{cf.} appendix \ref{vornoi_appendix}).\\

We assume that building a new substation (HV/MV or MV/LV) and laying kilometers of lines are both costly, and the network seeks a design that reflects cost and engineering trade-offs \cite{persoz_planification_1984}. 'Costly' refers here to a general cost concept, not only
the direct building cost (e.g., the electricity loss associated with a particular configuration). Having a node position dataset, we can connect the substation using a minimum spanning tree algorithm. The total length of such a network is determined by the density of nodes and the distances between them \cite{barthelemy_spatial_2022}. In real distribution networks, it is cost-effective to add loops to the tree structure: when the nodes are connected to the rest of the network through at least two paths, it is safer and more effective in terms of Joule losses \cite{crastan_reseaux_nodate2}. Nevertheless, the contribution of bridges' length to the network's total length is weak, and the tree-like structure serves as a suitable first proxy for estimating the overall spatial layout and material content of an MV grid \cite{abeysinghe_topological_2018}. An estimation of the contribution of loops to the overall length of the French distribution grid is made in \cite{emery_statistical_2025}, and shown to be negligible. Still, the second-order corrections to the total cable length estimates are given in appendix \ref{correction_appendix}.

We emphasize that the generated networks are geometric objects: nodes and edges carry no electrical attributes (impedances, capacities, loads), and the outputs are not intended for power-flow or operational analyses. The framework is an estimation tool for network extent, spatial layout, and material content. Throughout this paper, route length refers to the length of the physical paths (which may carry several cables), while cable length counts each cable individually; the conversion factor is calibrated in appendix \ref{correction_appendix}.

\begin{figure}[t]
    \centering
    \includegraphics[width=1\linewidth]{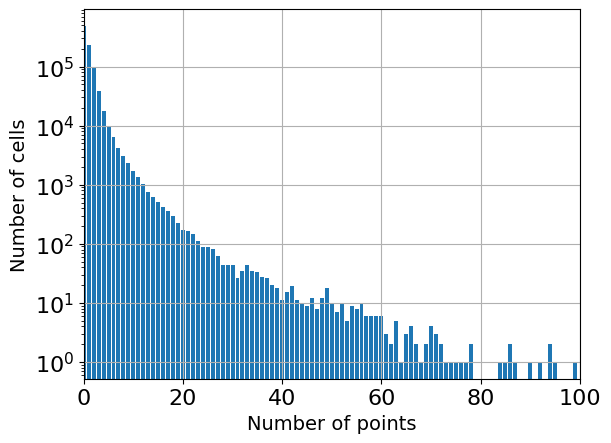}
    \caption{Histogram of the number of nodes in grid-cells.}
    \label{fig:histogram}
\end{figure}

\begin{figure}[t]
    \centering
    \includegraphics[width=1\linewidth]{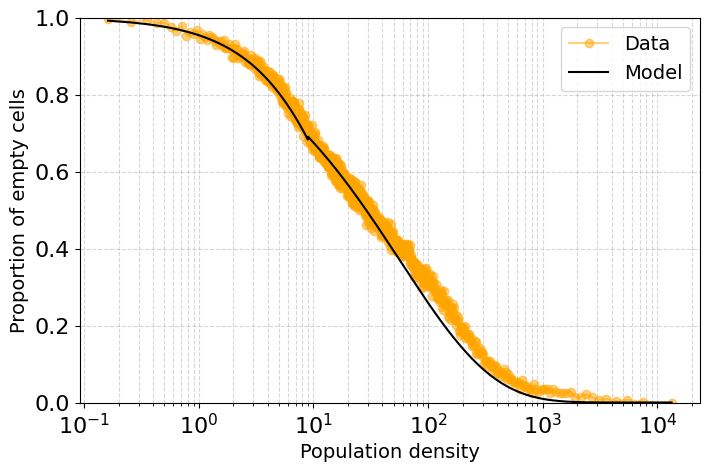}
    \caption{Proportion of empty cells in the data for all population density values.}
    \label{fig:prop_null}
\end{figure}

\section{Results}

This Results section is structured into three subsections, progressing from a local to a global perspective on the relationships between population, spatial organization, and the electricity grid. First, focusing on the French case, we compare the outputs of the proposed model—based on population density data at a 30-arc-second spatial resolution (approximately $0.59\,\text{km}^{2}$ at French latitude)—with the existing electricity network. The similarity between the generated and observed networks is assessed qualitatively through local-scale visual inspection and quantitatively using aggregated indicators at the global scale. The agreement between the model-generated network and empirical data confirms the relationship between network structure and the spatial distribution of population. Moreover, the model's construction enables this relationship to be consistently captured at both local and global scales.

Second, we apply this network generation process, calibrated on the French case, to other countries for which detailed local grid data are unavailable but aggregated network statistics are available. The analysis of these results enables to take a few steps toward the generalization of the relationship between electricity networks and population distribution, as well as an evaluation of the extent to which a model calibrated at the local scale in France can reproduce macroscopic quantities (eg., total network length) in other national contexts.

Finally, we investigate this relationship directly at the global scale by examining, among other variables, aggregated measures of territorial area and population (number of households) in relation to the total lengths of different voltage classes across a large number of countries.  This approach leads to the emergence of simple scaling laws that aim to capture some degree of universality at the macroscopic level.

\subsection{Generating the French MV network from sole population density}

\subsubsection{Visual comparison at the local scale}

\begin{figure*}
    \centering
    \includegraphics[width=1\linewidth]{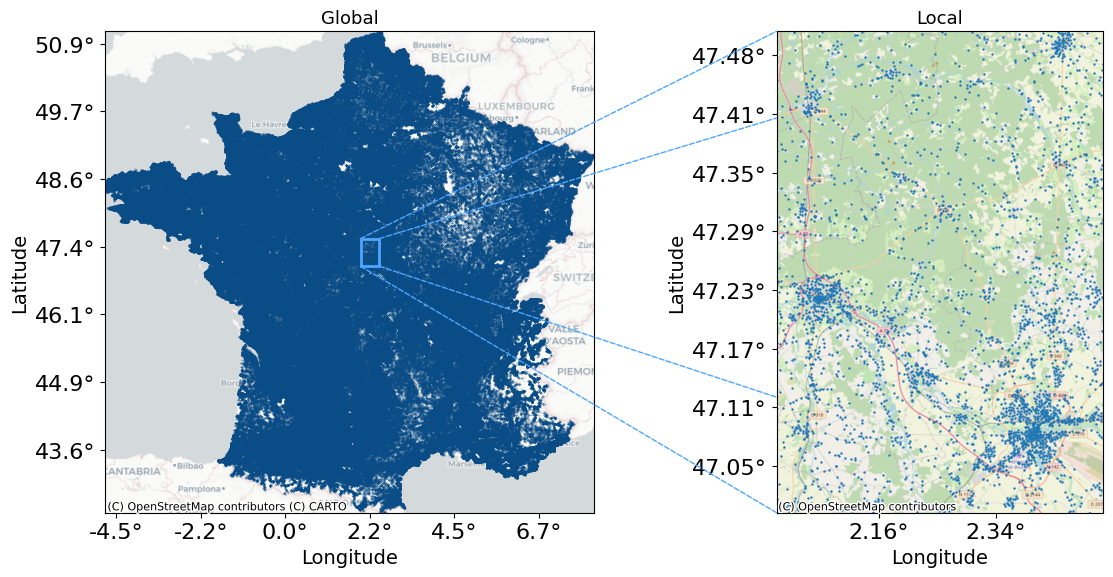}\\

    \includegraphics[width=\linewidth]{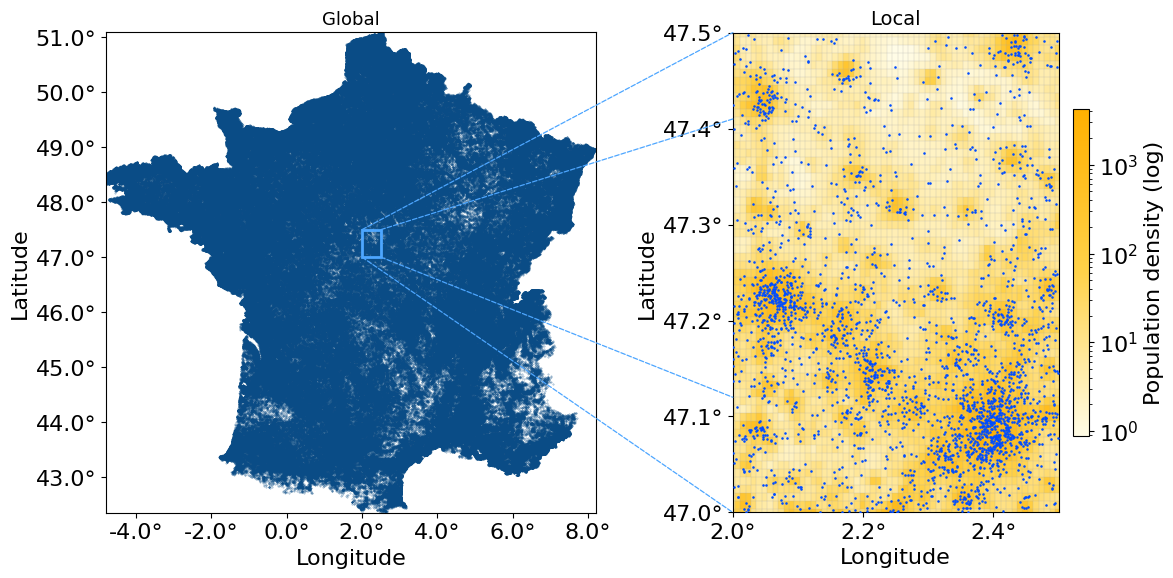}

    \caption{Node spatial distribution for the empirical dataset (top figure), and the model (bottom figure). The white holes in the top figures correspond to areas that are not managed by ENEDIS.}
    \label{fig:visualNode}
\end{figure*}

\Cref{fig:visualNode} shows the locations of the network nodes generated by the model together with the actual locations of nodes across the entire territory. A strong similarity is observed at both local and global scales.

The white areas in the map shown in the upper part of \Cref{fig:visualNode} (empirical data) correspond to territories for which no substation data are available from any distribution operator (ENEDIS or local distribution companies). These residual gaps cover only about 2.5\% of metropolitan France in surface and host merely 0.28\% of its population, as they are essentially rural. This artifact is therefore negligible.

\begin{figure*}
\begin{subfigure}{0.49\linewidth}
        \centering
    \includegraphics[width=1\linewidth]{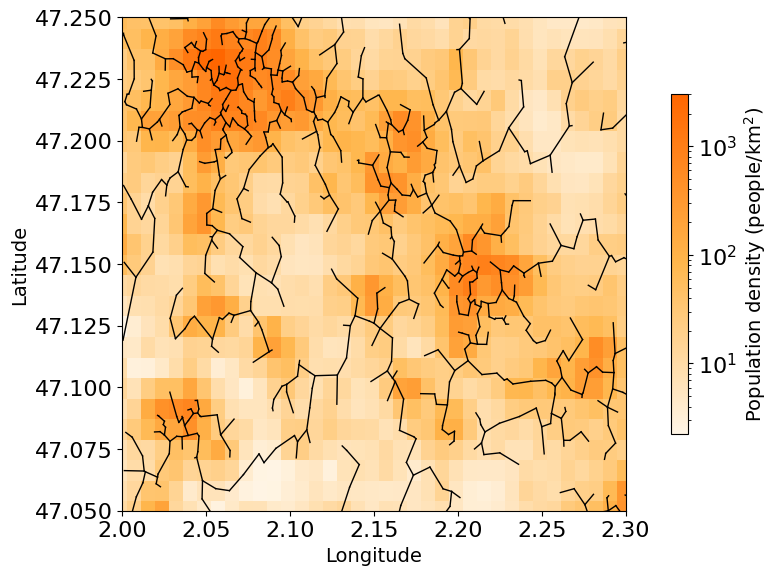}
    \caption{}
    \label{fig:visualNetworkMODEL}
\end{subfigure}
\begin{subfigure}{0.49\linewidth}
        \centering
    \includegraphics[width=0.85\linewidth]{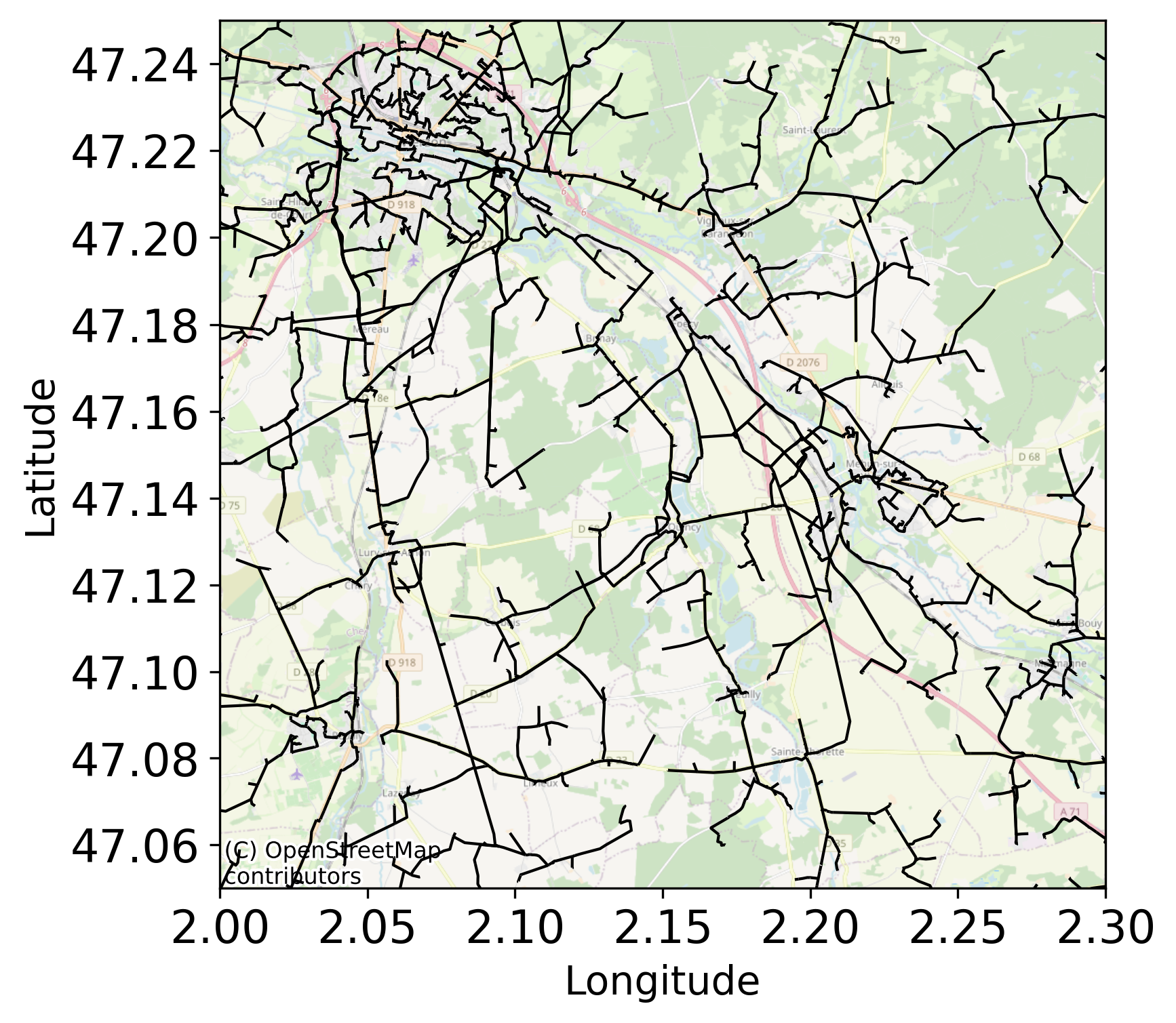}
    \caption{}
    \label{fig:visualNetworkDATA}
\end{subfigure}
    \caption{(a) Real population density data in France (orange field) with minimum spanning tree edges (black lines). (b) Real-world distribution grid layout in the corresponding area, with blue nodes representing MV/LV substations and black edges indicating MV lines.}
    \label{fig:visualNetwork}
\end{figure*}

\Cref{fig:visualNetwork} no longer displays individual nodes but instead shows the full set of edges generated by the model, together with a comparison to the actual line drawings provided by the grid operator ENEDIS. The empirical MV network is close to a tree, even though some loops exists. This empirical observation illustrates our modelling choice based on minimum spanning trees. The model reproduces the dominant skeleton of the real network by construction, the meaningful comparison is on the spatial placement and density of edges.

By contrast, the edges are not necessarily spatially positioned with the highest precision. In particular, we observe a specific effect induced by the road network: a fraction of substations and the majority of power lines are aligned along roads. Population density can partially account for this effect, as roads tend to pass through cities and villages with higher population density, but it fails to do so outside inhabited areas. Thus, the main source of error is associated with the low-density areas, which contain a low number of substations. Still, the validation of our model suggests that this error does not affect the aggregated network length. Regarding the realistic road alignment, this effect is hard to address computationally. Indeed, we could imagine a multi-layer model that draws a synthetic road network that correlates with the distribution grid topology \textit{a posteriori}. We choose to rely on a simple correction heuristic, detailed in Appendix \ref{correction_appendix}.

Another, more subtle, topological feature of the real network is the presence of \textit{bridges}, that is, connections between nodes that are already indirectly connected through two branches of the tree. This addition of links relative to a purely tree-like topology provides a degree of resilience to the real network. These bridges are visible through the loops present in the right-hand panel of \Cref{fig:visualNetwork} but absent in the left-hand one. Accounting for this specific feature would require adapting the node-connection algorithm (discussed in appendix \ref{correction_appendix}).

In these respects, it is important to emphasize that the generated output represents a “typical” network corresponding to the population distribution. This network approximates the locations of substations and connectivity patterns, but it does not aim to predict whether an actual cable passes through given spatial coordinates.

The model output cannot resolve sub-cell features. Above this resolution, aggregated indicators are expected to be more and more reliable as the length scale increases. Even though the generated network reproduces approximations of grid topology and node spatialization, highly local effects (eg., those induced by specific basement layouts) are averaged out at larger spatial scales. This is developed in appendix \ref{multiscale}.

Finally, unlike ENEDIS, which as previously mentioned provides two distinct datasets (nodes and lines drawings), this model produces a fully consistent network composed of both nodes and edges, with a well-defined topology.

\subsubsection{Quantitative features at the country scale}

The underlying hypothesis of this work—that the electricity network is determined, or at least strongly correlated, with the spatial distribution of population—appears to be confirmed at the local scale in the French case, where the network seems to follow the population density field. A natural question is whether the apparent effectiveness of the proposed modeling approach at the local scale is preserved at the global scale when considering aggregated network characteristics. In particular, we focus on two essential quantities: the total length of the network and the number of substations.

This is by no means guaranteed \textit{a priori}. Regarding the number of substations, the node-density relation $f(\rho)$ is calibrated on the French data. Recovering the national substation count is therefore a consistency check rather than a validation. This check is nonetheless non-trivial, the total number of substations is not imposed as a direct constraint, and the parameterization concerns only the expected number of substations as a function of the population density within a cell, which is itself modeled using a weakly parameterized power-law relationship. Concerning the second feature, the total network length is not incorporated into the sampling parameterization at all; it emerges from the tree structure applied to stochastically sampled nodes, and its comparison with the empirical value constitutes a validation of the connection step. We note, however, that this comparison is not a fully independent validation: the route-to-cable and loop correction factors ($\gamma_2$, $\gamma_3$) used to convert the modeled route length into cable length are themselves partly calibrated on French ENEDIS data (appendix \ref{correction_appendix}). The fully independent test of the framework as a whole is provided by its application to 35 other countries (Sec. \ref{subsectionB}).

\Cref{tab:FrenchIndicatorComp} presents a comparison between the values of these quantities computed from the generated model and the experimental data obtained for the actual network. While the accuracy of the estimated total number of nodes ($88\%$), as a consistency check, is not particularly surprising, the accuracy of the total network length ($94\%$) highlights a key strength of the model: at the global scale, the local discrepancies in line routing observed in \Cref{fig:visualNetwork} have no impact or effectively compensate for one another (see appendix \ref{multiscale}). That is, the difference in total line length within a given area (e.g., between Figures \ref{fig:visualNetworkMODEL} and \ref{fig:visualNetworkDATA}) is consistently very small, or the over- and under-estimation errors offset each other when aggregated over the entire French territory. In both cases, the generated network can therefore be interpreted as a representative or “typical” network that preserves both the local visual structure and the values of global network characteristics.

\begin{table}[t]
    \centering
    \begin{tabular}{|c|c|c|c|}
    \hline
        & $N_{stat}$ ($10^{5}$) & $L_{MV}^{MST}$ ($10^{5}$ km) & $L_{MV}^{cable}$ ($10^{5}$ km) \\
        \hline
        Literature & 7.87 & n.d. & 6.45 \\
        \hline
        Model & 8.79 &  3.94 & 6.06\\
        \hline
        Error & 12\% & n.d. & 6\%\\
        \hline
    \end{tabular}
    \caption{{\bf French aggregated indicator comparison} The total number of substations and the total cable length computed from the generated network are close to the values reported in the literature (\textit{cf} appendix \ref{correction_appendix} for details on the error corrections).}
    \label{tab:FrenchIndicatorComp}
\end{table}

The model constructed for the French case thus establishes a strong link between population distribution and the electricity network at both local and global scales. This suggests that a simple local population–network relationship is sufficient to iteratively (“cell by cell”) construct a global network that approximates the real one. Beyond the network generation process itself, which may be valuable for future research or prospective studies on network deployment, this result indicates that, despite (i) the geographical, social, and historical diversity of the various territories composing France and (ii) the presence of actors and potential constraints operating at the national rather than the local level, a single local relationship is, on its own, sufficient to characterize key properties of the electrical network across all local areas as well as at the national scale.

\subsection{Results for other countries}\label{subsectionB}

This correlation law between population density and the density of electrical substations, therefore, appears to provide an effective description of the current French power grid. A natural subsequent question is whether this descriptive capability remains valid at other points in time or in other countries. Since no data from these countries enters the calibration, this cross-country comparison constitutes the independent validation of the framework. Regarding spatial extension, the following section presents a first comparison of the model results across 35 countries. In principle, the model output would allow comparisons at the local scale and analysis of differences in network topology and substation locations between France and other countries. Unfortunately, the lack of access to local-scale data on electrical networks in other countries has prevented such a comparison.

Nevertheless, as in the French case, this approach still makes it possible to compute the global characteristics of the power grid (total number of substations and total network length) from the model results, and to compare them with values reported in "grey" literature for the different countries, which are available for all selected cases. These quantities must, however, be interpreted with caution, as they are sometimes derived from estimates based on strong and various underlying assumptions. For instance, the reference years differ across countries, and the voltage boundaries defining the MV class vary between national systems. In order to ensure that this comparison is as meaningful as possible, we selected countries for which the relevant data were available and that exhibit an electricity access rate similar to that of France (access to electricity for $100\%$ of the population as defined by the World Bank \cite{worldbank2023electrification}). This factor, therefore, cannot account for potential differences relative to the French case, which must instead be sought in the manner in which the networks are organized.

\begin{figure}[t]
    \centering
    \includegraphics[width=\linewidth]{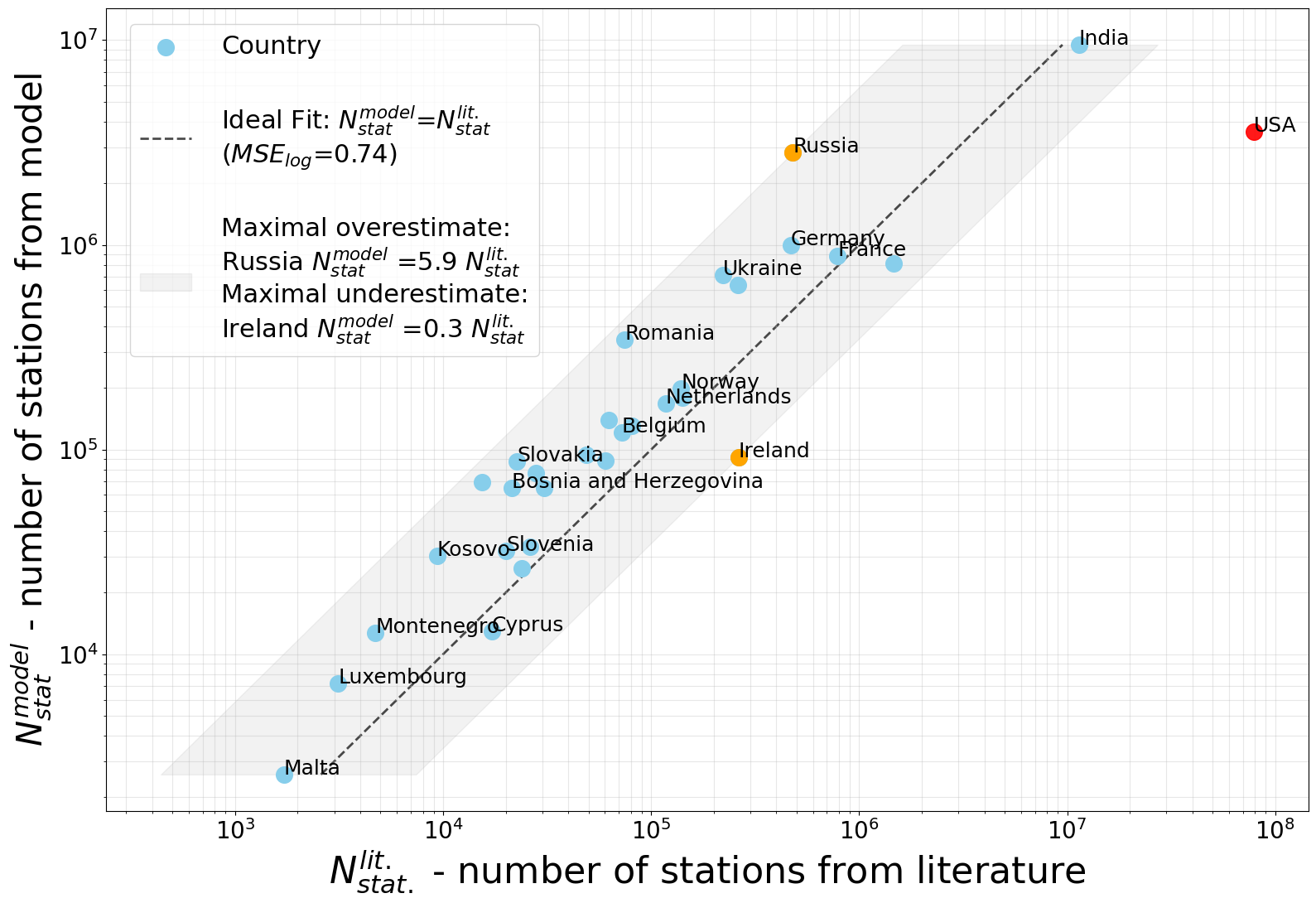}
    \caption{Comparison of total $N_{stations}$ value computed on model output and estimation from literature.}
    \label{fig:modelNstat}
\end{figure}

\begin{figure}[t]
    \centering
    \includegraphics[width=\linewidth]{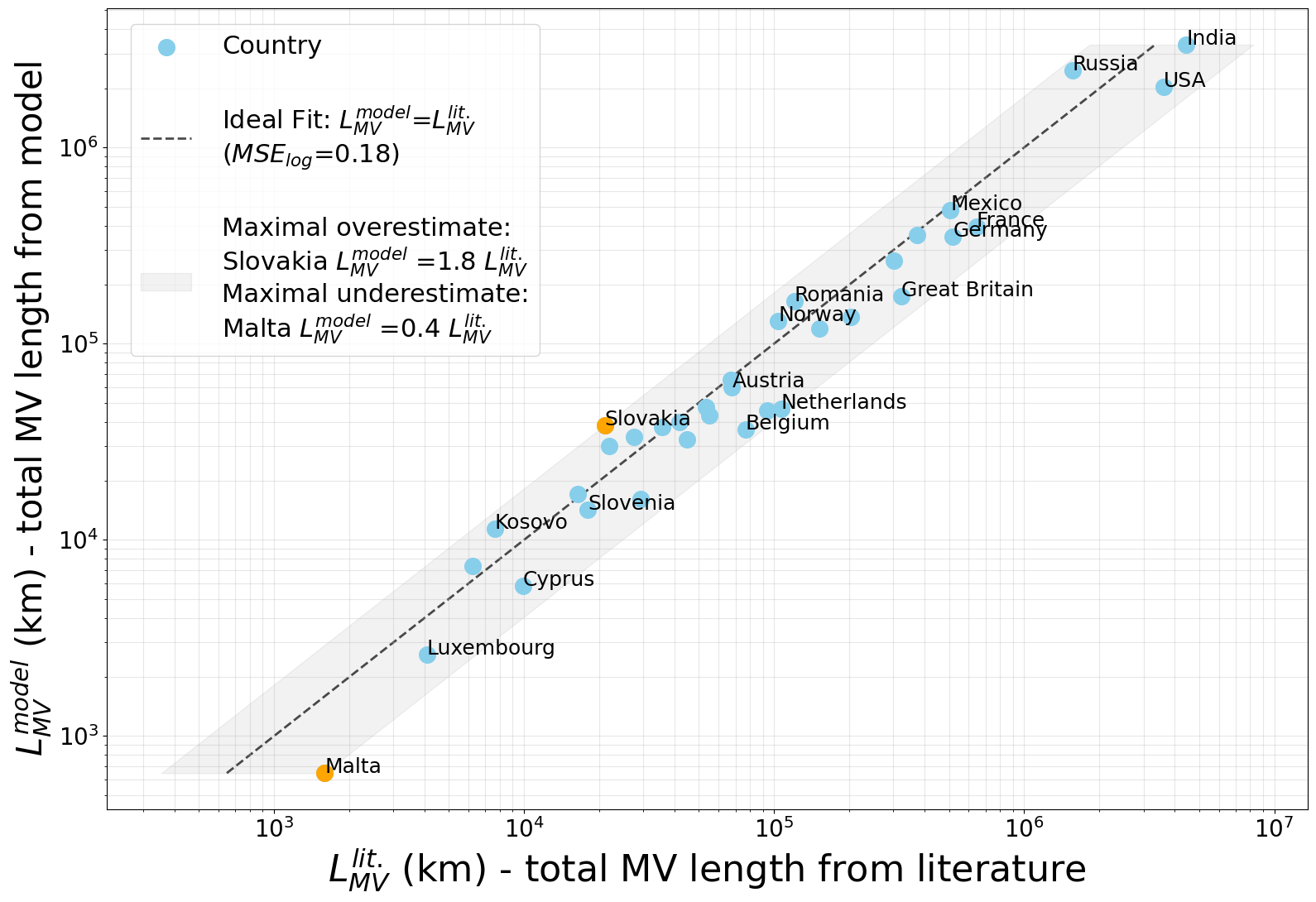}
    \caption{Comparison of total $L_{MV}$ value computed on model output and estimation from literature. It represents the total MV network lengths reported in the literature and those computed by the model. As it is often unclear whether the values reported in the literature refer to route length or cable length, and how the latter is calculated, we chose in this case to use as model output the direct sum of segment lengths, without applying any correction factors (\textit{cf} appendix \ref{correction_appendix}).}
    \label{fig:modelMV}
\end{figure}

Figures \ref{fig:modelNstat} and \ref{fig:modelMV} present a comparison between the total number of substations and the total network length computed from the model outputs—generated using each country’s local population density and a model calibrated on the French case ($y$-axis)—and the corresponding values reported in the "grey" literature ($x$-axis). These figures should be interpreted as follows: each point represents a country; the closer a point lies to the line $y=x$, the closer the model-derived value is to the value reported in the literature. Points above this line indicate model overestimation, whereas points below it indicate underestimation.

The substantial disparity in magnitudes across countries makes a log–log representation visually more meaningful, but it also considerably weakens the interpretability of the $R^{2}$ metric, which is almost entirely driven by the extreme values in the distribution tail (e.g., India, the United States, or Russia) that dominate the remaining data points. In this context, it is therefore more appropriate to consider the mean squared logarithmic error ($MSE_\text{log}$), which penalizes relative deviations and is independent of the order of magnitude of the values. This indicator is computed as follows :

\begin{equation}\label{MSElog}
MSE_\text{log} = \frac{1}{N_\text{countries}}\sum_{i\in \text{Countries}} \log\Big(\frac{y_{\text{model,}i}}{y_{\text{lit.,}i}}\Big)^{2}.
\end{equation}

With respect to the number of substations, Figure \ref{fig:modelNstat} shows that the model exhibits a substantial logarithmic error. Both a significant bias—manifested as a tendency of the model to overestimate the number of substations (points lying on average above the line $y=x$)—and considerable dispersion are observed. The United States of America, which was excluded from the dataset used for the calculations, shows a discrepancy of two orders of magnitude relative to the model output, highlighting the non-universal nature of the relationship employed.

In fact, there appear to be “two basic models for distribution systems, the European and the American, with combinations of the two employed by various countries. These models use different mixtures of distribution transformers, with the European system having a higher concentration of larger-capacity, higher-voltage MV transformers and fewer small units, while the American system has a large number of small units located closer to the customer.”\cite{TransformerReport}. This design difference, which translates into a substantial discrepancy in the number of substations at the macro level, necessarily also affects the network's local configuration. The population–substation relationship is therefore architecture-dependent: the calibrated relation $f(\rho)$ encodes not only the effect of population density but also the transformer granularity specific to a France-style system, and cannot be expected to transfer to systems built on a different design. It can nevertheless be hypothesized that countries for which the model closely matches the literature values actually exhibit a distribution system design similar to that of the French case.

With respect to the total medium-voltage network length, Figure \ref{fig:modelMV} shows a much closer agreement between the model results and the literature values. Only a small bias is observed, with a slight tendency toward underestimation, and the points are highly concentrated around the line $y=x$. This finding is notable given the disparities in $N_{stat}$ and, consequently, in the local network configuration. It appears that, despite differences in the topologies of foreign networks, the model calibrated on the French case provides a reasonable aggregate estimate of the total length of the medium-voltage network. This may suggest that the nature and number of transformers have relatively little impact on the total network length, which instead appears to correlate strongly with the population's spatial distribution. This contrast between the two metrics can be understood as follows. The substation count directly reflects the transformer granularity, which is an architectural choice. The network length, by contrast, is governed by the spatial extent of the demand to be covered: the MV network must reach every inhabited area regardless of how many transformer nodes subdivide it. In American-style systems, we can hypothesize that numerous small transformers are predominantly placed along existing routes, thereby increasing the node count while leaving the route length essentially unchanged. More generally, the length of a spanning tree over $N$ nodes distributed in an area $A$ scales as $\sqrt{N A}$ \cite{barthelemy_spatial_2022}, so that even variations in node count by a factor $k$ would only affect the length by $\sqrt{k}$ and much less when the additional nodes are spatially correlated with existing lines. Total network length therefore correlates strongly with the population's spatial distribution while remaining less sensitive to transformer count.
\subsection{Macro-level indicators estimation}\label{subsectionC}

The relationship between population—sometimes measured by the number of households—and the total length of electrical network cables is not new and has been highlighted in the literature \cite{kalt_global_2021, zvoleff_impact_2009, arderne_predictive_2020, bettencourt_growth_2007, kuhnert_scaling_2006}. However, there is no consensus on how to characterize this relationship. Kalt \textit{et al.} employ a linear regression with respect to the number of households connected to electricity and the country’s electricity consumption to estimate the total length of the distribution network \cite{kalt_global_2021}. Other authors, such as Kuhnert and Bettencourt, observe an empirical power-law relationship between the length of low-voltage cables and population in densely urbanized areas of Germany, with an exponent ranging from 0.81 to 0.92 \cite{bettencourt_growth_2007, kuhnert_scaling_2006}.

It is noteworthy that in these previous formulations, the spatial dimension (geographic dispersion of the population) does not appear to affect the total cable length. This naturally raises questions, as it seems intuitive that the same population spread over a larger territory would require a more extensive network, and thus a greater total cable length. In the case of Kuhnert and Bettencourt, the absence of a spatial effect may be explained by the restriction of their study to dense urban areas, thereby limiting its spatial extent. Nevertheless, it seems reasonable to observe a sublinear scaling law in cities, since urban networks, regardless of their nature, exhibit fractal characteristics \cite{bettencourt_origins_2013}, capturing scale effects that a linear relationship cannot. Indeed, a clearly sublinear exponent is also observed in the model calibration for urban areas (see Section \ref{sec:Model}).

This local-scale power-law relationship, as shown in the previous section, enables reasonable aggregate estimates of the length of the medium-voltage network. Given the existence of general empirical laws governing how populations occupy space, whether in urban or rural areas \cite{cattaneo_economic_2022, emery_power_2025}, the local power-law relationship may be expected to leave a macroscopic signature: aggregating a density-dependent local law over a territory whose population follows regular spatial distribution patterns could yield simple scaling relations between total network length, total population, and land area. The two approaches are therefore complementary: the bottom-up model establishes the mechanism and provides spatially explicit outputs, while the top-down regressions test its emergent macroscopic signature and offer a shortcut requiring only aggregate indicators. The following section presents multivariable power-law regressions incorporating population (measured by the total number of households) and national surface area, along with a discussion of the impact of other variables such as electricity consumption.

\subsubsection{Power-law scaling of the number of substation - $N_{stat}$}

Consistent with the observations of the previous section, $N_{stat}$, which was poorly captured by the model due to the existence of multiple possible network architectures, is also relatively poorly described by a multivariable power-law regression involving population, national surface area, and electricity consumption. The regression yielding the best performance is: 
\begin{equation}    
N_{stat} = 0.07\,Pop^{0.75}\,Area^{0.22}.
\end{equation}
However, it is characterized by unsatisfactory values of
$MSE_\text{log}=0.32$ . This confirms that, without specifying the underlying network architecture, it is difficult to relate population density to substation density at the global scale.

\subsubsection{Powerlaw-scaling of the medium-voltage length - $L_{MV}$}

\begin{figure}[t]
    \centering
    \includegraphics[width=\linewidth]{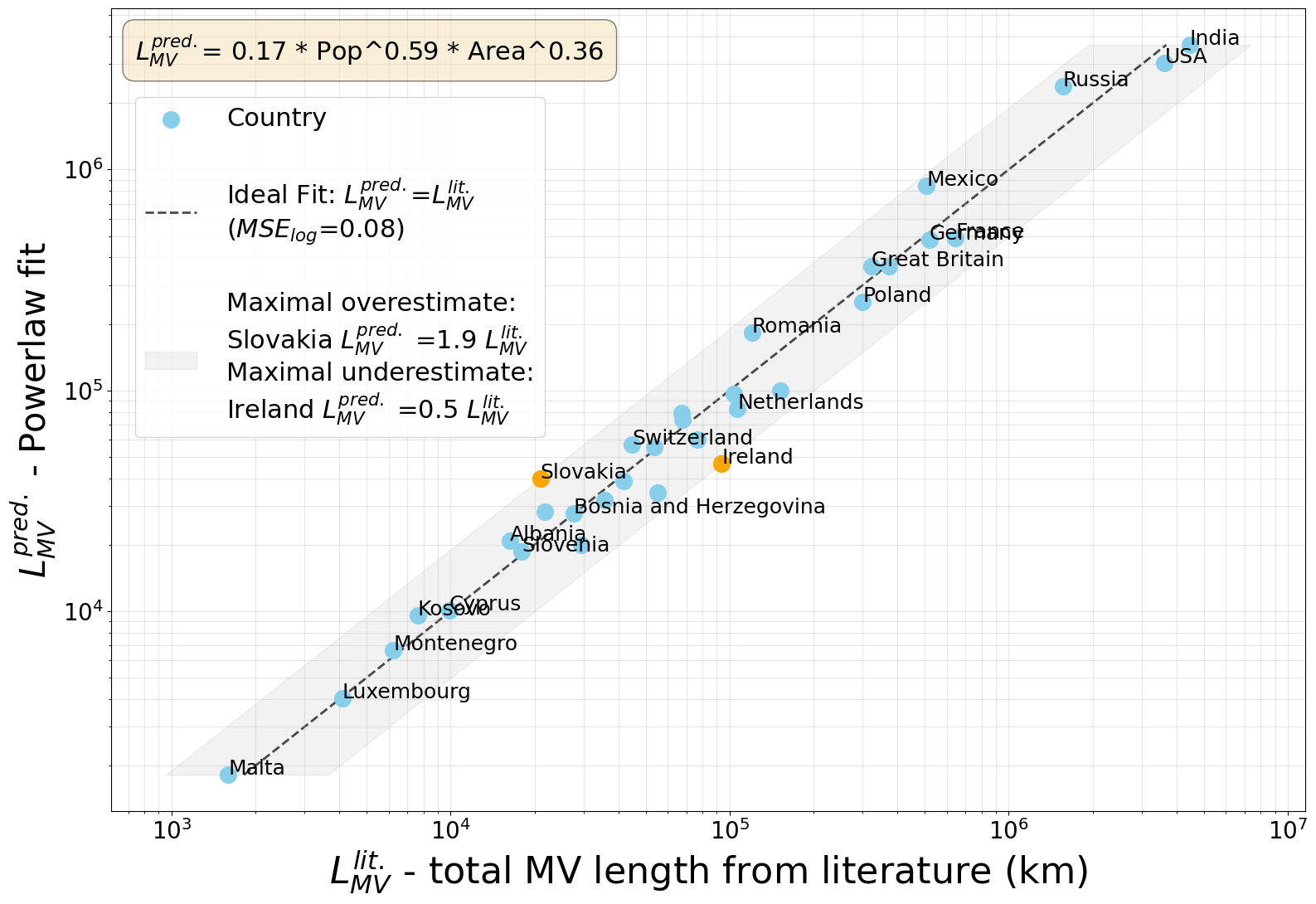}
    \caption{Estimation of total $L_{MV}$ value using powerlaw relationship on the country's total population and area.}
    \label{fig:powerlawMV}
\end{figure}

By contrast, and in line with the model's results, the medium-voltage network length is particularly well described by power-law relationships involving global variables. The equation providing the best fit is the following: 
\begin{equation}
    L_{MV} = 0.05\,Pop^{0.6}\,Area^{0.35}\,Conso^{0.13},
\end{equation}
with
$MSE_\text{log}=0.076$. However, an even simpler formulation that does not include electricity consumption yields estimates of nearly identical accuracy:
\begin{equation}\label{length_pop_area}
L_{MV} = 0.17\,Pop^{0.59}\,Area^{0.36},
\end{equation}
with
$MSE_\text{log}=0.081$. Figure \ref{fig:powerlawMV} compares the results of this power-law relationship ($y$-axis) with the values reported in the literature ($x$-axis) for the countries considered. A correlation is observed that is even stronger than that obtained from the model-based results. The relatively minor impact of explicitly accounting for electricity consumption is a noteworthy finding, which is discussed further in the next section.

\subsubsection{Power-law scaling of the low-voltage length - $L_{LV}$}

Although our model focuses on the medium-voltage network—primarily due to the availability of local-scale data only for this voltage class—the works of Kalt \textit{et al.} and Bettencourt suggest that the relationship between population and cable length may also hold at the low-voltage level \cite{kalt_global_2021, bettencourt_growth_2007}. The LV layer is not spatially constructed by our model, but following this line of reasoning, we applied the same power-law fitting procedure. The equation providing the best fit is :
\begin{equation}
    L_{LV} = 0.4\,Pop^{0.66}\,Area^{0.22}\, Conso^{0.26},
\end{equation}
with
$ MSE_\text{log}= 0.14$. These results suggest that, in contrast to $L_{MV}$, the low-voltage network length depends less strongly on the total surface area of the country. This observation is consistent with intuition: the low-voltage network represents the “last layer” of the distribution tree, located close to where people actually reside, regardless of how geographically dispersed population clusters may be.

\begin{figure}[t]
    \centering
    \includegraphics[width=\linewidth]{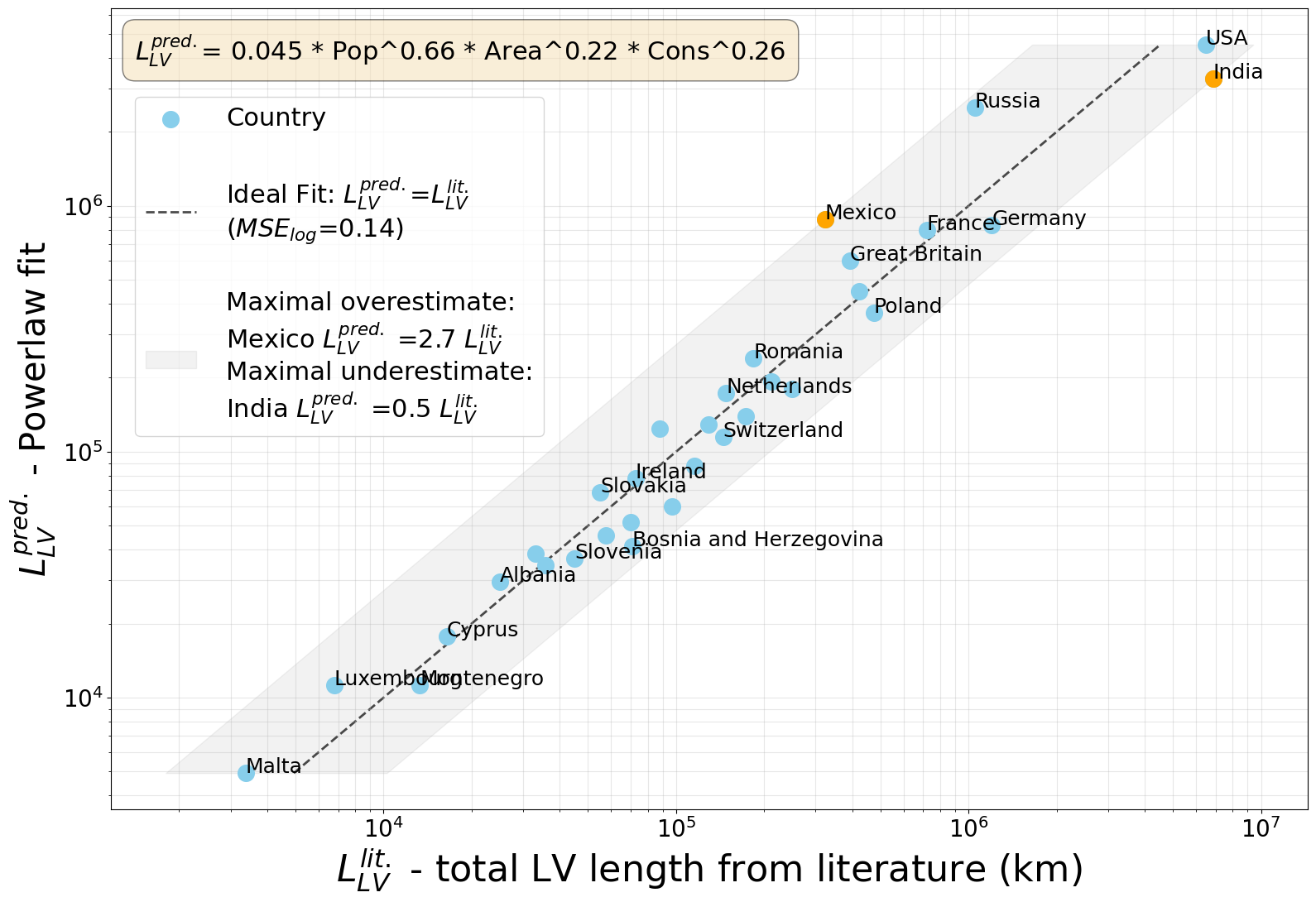}
    \caption{Estimation of total $L_{LV}$ value using powerlaw relationship on the country's total population and area.}
    \label{fig:powerlawLV}
\end{figure}

\subsubsection{Correlation between $L_{LV}$ and $L_{MV}$}

\begin{figure}[t]
    \centering
    \includegraphics[width=\linewidth]{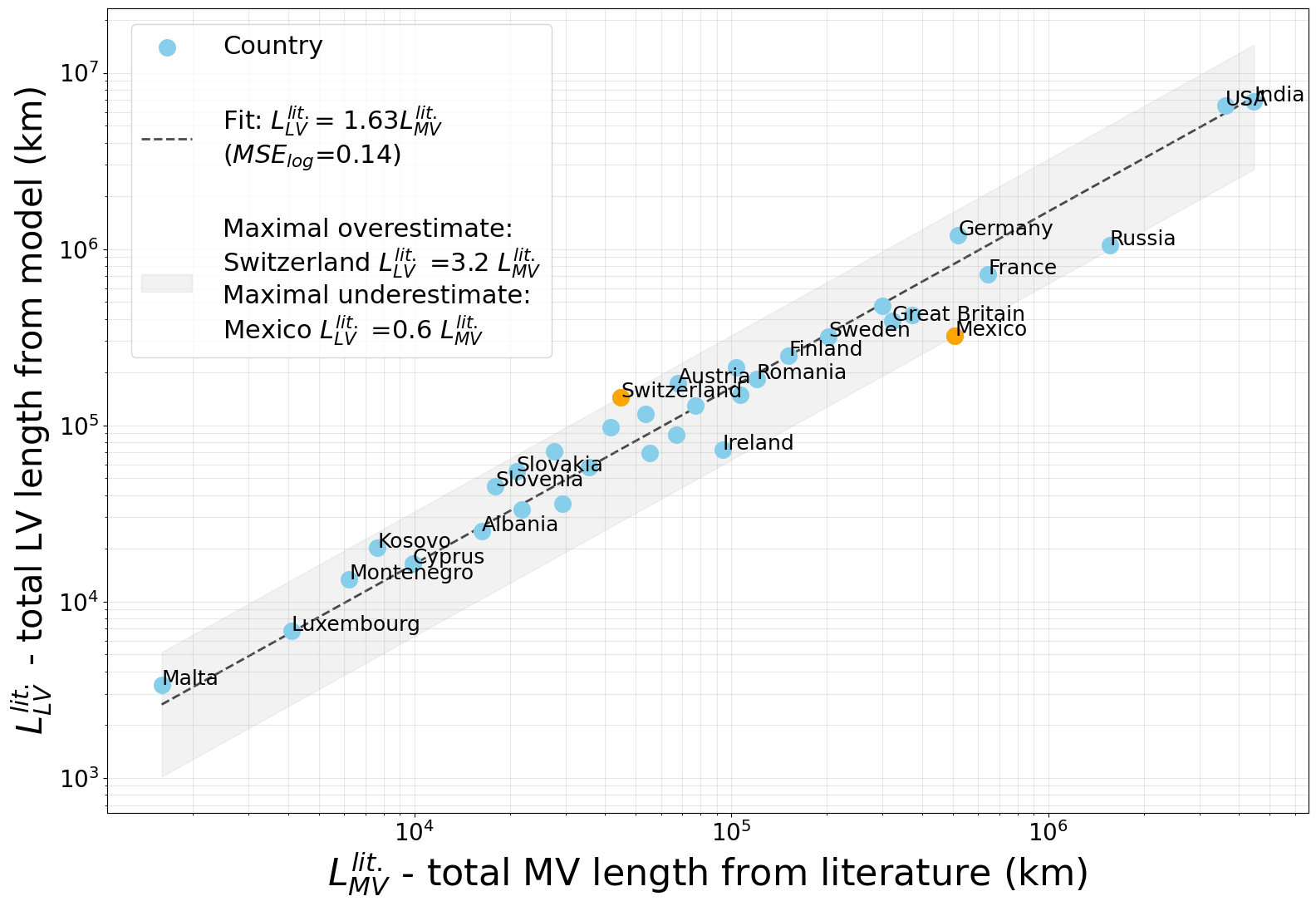}
    \caption{Estimation of total $L_{LV}$ value using linear regression on $L_{MV}$ value.}
    \label{fig:LV_MV}
\end{figure}

A fairly strong correlation is also observed between the lengths of low-voltage and medium-voltage cables. Kalt et \textit{al.} assume that low-voltage cables account for $63\%$ of the total length of distribution network cables. By fitting the low-voltage length as a function of the medium-voltage length, we obtain a coefficient of $1.63$ ($MSE\text{log}=0.14$), indicating that the low-voltage length is $1.63$ times greater than the medium-voltage length (Figure \ref{fig:LV_MV}). This corresponds to a share of $\frac{1.63}{1+1.63}=62\%$, which is close to the value reported by Kalt et \textit{al}.

\begin{figure*}[t]
    \centering
    \includegraphics[width=0.8\linewidth]{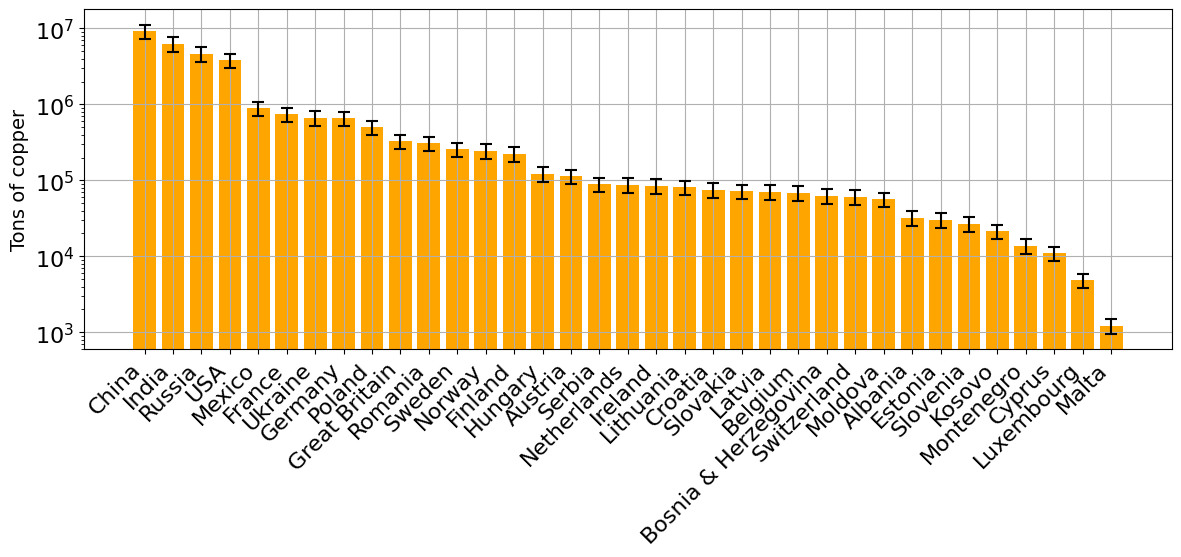}
    \caption{Approximation of the total amount of the copper invested in MV grid for various countries using corrected length values from the network algorithm. The upper bound is reached if there are only aerial lines, and the lower bound is reached if there are only underground lines.}
    \label{fig:copper}
\end{figure*}
\subsection{Material investment scaling with network length}\label{subsectionD}

Having a heuristic for the total cable length enables an inventory of the material invested. While the following analysis focuses on copper, the methodology remains applicable to other material types. Using the Surfer database from ADEME \cite{surfer}, we can obtain consensus values for material intensities for power grid lines. These values are estimated from various sources for a French type of MV line, and are expressed in tons per kilometer of cable. The copper intensity of MV line is $\xi_a \simeq 1.49$ t/km for aerial lines and $\xi_u \simeq 0.96$ t/km for underground lines. Once the network is constructed, multiplying the total MV length by $\xi_u$ (respectively multiplying by $\xi_a$) yields a lower bound (respectively an upper bound) of the total amount of copper invested in lines. These estimates are presented in \Cref{fig:copper}.\\

An even more straightforward alternative consists in leveraging macro-level indicators to derive material estimates. In this regard, \Cref{length_pop_area} could be transformed into a proxy for estimating copper mass:
\begin{equation}
m_{C_\text{u}}^{MV} = 0.17\,Pop^{0.59}\,Area^{0.36} \times (\xi_u \tau + \xi_a (1-\tau))
\end{equation}
where $\tau$ is the fraction of underground cabling. This parameter is inherently difficult to estimate, as it fluctuates significantly across jurisdictions according to regulatory frameworks, aesthetic preferences, and economic constraints. This represents another dimension in which different national design traditions can produce major architectural differences. While this methodology remains applicable to the LV layer, the predominance of underground infrastructure at this level renders the uncertainties surrounding $\tau$ particularly critical.

\section{Discussion \& Conclusion}

We have developed a simple modeling approach that simulates the spatial distribution of medium-voltage substations from high-resolution population density fields. From this node distribution, we derive a first order approximation of the associated MV network topology, yielding a refined estimate of the total cable length. The WorldPop database provides population density data for each country and year between 2000 and 2020, and our method has been benchmarked against reported aggregate data for 35 countries with diverse territorial characteristics.

Inputting high-resolution population data in our network-generation algorithm enables us to derive a first-order spatial allocation of likely in-use copper stocks. This spatial information supports targeted material flow analysis and territorial planning, enabling the identification of where metallic resources are likely concentrated within the distribution grid. These results are particularly relevant to global electrification scenarios and infrastructure planning. For instance, the work of Vidal \textit{et al.} on future energy infrastructure requirements highlights the need for robust methods to estimate distribution network expansion \cite{vidal_modelling_2021}. Such methods can be applied in emerging economies where detailed grid records are scarce, enabling forecasts of infrastructure and material needs accompanying electrification.

This local-scale framework also serves as the foundation for analyzing global scaling properties of electricity distribution networks. We conducted a multivariate analysis to estimate properties of the MV/LV grid from macro-level indicators. A remarkable finding is that electricity consumption can be omitted from the scaling relationships for medium-voltage network length without significantly degrading predictive accuracy. This suggests that, at the national scale, the spatial distribution of population and the territorial extent capture the primary drivers of network extent, while the consumption effect adds limited explanatory value. This observation may reflect an emergent property of distribution networks: once population distribution and territorial extent are specified, the MV network topology is largely determined by the spatial optimization required to connect distributed loads. However, our results demonstrate that this does not apply to the low-voltage layer, where the contribution of consumption remains significant. This observation is consistent with expectations: at this scale, the infrastructure serving households is inherently localized, and meeting power capacity requirements constitutes the primary design constraint.

Taken together, these results demonstrate that useful representation of distribution grid structures can be generated solely from population density, and that aggregate network characteristics follow predictable scaling relationships with population and area. Because cable length governs the material content of a grid, these results can be translated into estimates of copper and aluminum demand for network deployment, as demonstrated in Section \ref{subsectionD}, making them particularly actionable for resource planning under electrification scenarios. 

Some important limitations need to be highlighted. The model represents only the spatial extent of grids (no technical/engineering characteristics involved) and is calibrated for mature, fully electrified MV distribution systems with European-style architecture. In particular, the population–substation relationship is architecture-dependent and does not transfer to American-style systems, whereas the total network length appears robust to this design difference (Sec. \ref{subsectionB}). Population density is thus a general driver of network extent, but not, on its own, of the substation count. To apply it to broader systems requires extra caution. Still, these findings provide new multi-scale tools for characterizing electrical infrastructure and reveal that population spatial distribution is a key determinant of distribution grid extent. From this understanding, these tools can be generalized in many ways. 

First, the network-generative process can be extended to the LV layer, branching the MV/LV substations down to end consumers. In this direction, high-resolution topographic databases such as the French BD TOPO \cite{BDTOPO}, which provides precise descriptions of buildings and infrastructure across the national territory, could help calibrate the spatial distribution of end-user connection points. In the end, this would yield a complete, end-to-end representation of the distribution network and refine material‑requirement estimates, particularly for the copper‑intensive last‑kilometer segments. 

Second, regarding temporal extension, the model could be adapted to population density time series to estimate the temporal evolution of the French grid. However, this approach relies on the assumption that the identified relationship was also valid in the past, an assumption that cannot be verified without sufficiently detailed historical data. It could nonetheless yield useful results for long-term studies—for instance, to describe the historical deployment of the French grid and to anticipate forthcoming replacement waves \cite{bihan_modeling_2025}—albeit subject to a strong underlying assumption. From a prospective perspective, the model appears considerably more suitable, as the current grid configuration likely reflects a degree of maturity and optimality that may persist into the future, thereby maintaining the descriptive power of the relationship. It is nevertheless important to emphasize the potential for major decentralization of electricity generation driven by the expansion of renewable energy sources, which could affect the distribution network and alter the observed empirical relationship.

Third, we can note that population density fields could serve as a common generation parameter for other infrastructure networks (water distribution, gas pipelines, or telecommunications, for instance) whose deployment logic is likewise driven by the spatial distribution of demand. Then, adapting the substation‑sampling and network‑construction steps to the specific technological constraints of each system would open the way to multi‑layer infrastructure modeling and to extended analyses of material needs.

\vspace{1cm}

\textbf{Acknowledgements}---We would like to acknowledge interesting discussions with Charley Presigny, Sourin Chatterjee and Seddik Yassine Abdelouadoud. \\

\textbf{Author contributions}---JH supervised this work. EE and JLB performed the investigation, formal analysis, software development, initial draft writing, final writing, review, and editing equally. All authors contributed to the study’s conceptualization and validation.\\

\textbf{Conflict of interest}---The authors have no conflict of interest to declare.\\

\textbf{Data availability}---The data generated by the authors for this study are openly available in the Zenodo repository at \url{https://doi.org/10.5281/zenodo.18634548}, together with the code that produces them.\\ 

\textbf{Code availability}---The code developed for this paper is available in the same Zenodo repository, at \url{https://doi.org/10.5281/zenodo.18634548}.\\

\textbf{Funding}---This work has benefited from a government grant managed by the Agence Nationale de la Recherche under the France 2030 program, reference ANR-22-PERE-0003.

\begin{appendices}

\section{Voronoi slicing}
\label{vornoi_appendix}

\begin{figure}[h]
    \centering
    \includegraphics[width=1\linewidth]{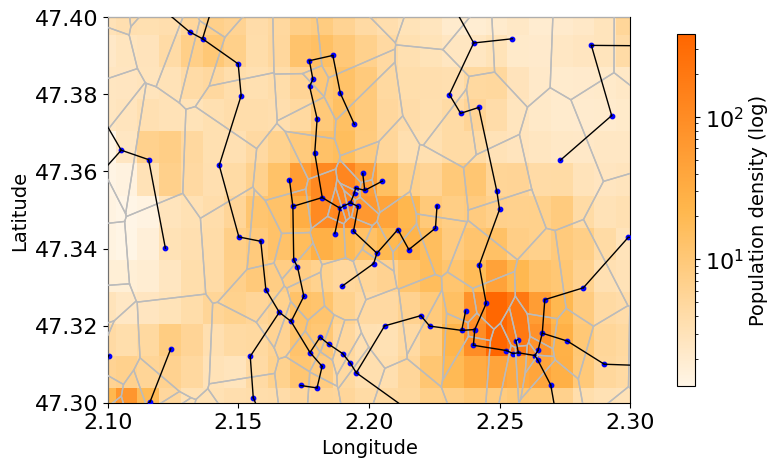}
    \caption{ Real population density data in France (orange) with simulated nodes (blue), minimum spanning tree edges (black), and computed Voronoi diagram (grey).}
    \label{fig:voronoi_mst}
\end{figure}

\begin{figure}[h]
    \centering
    \includegraphics[width=1\linewidth]{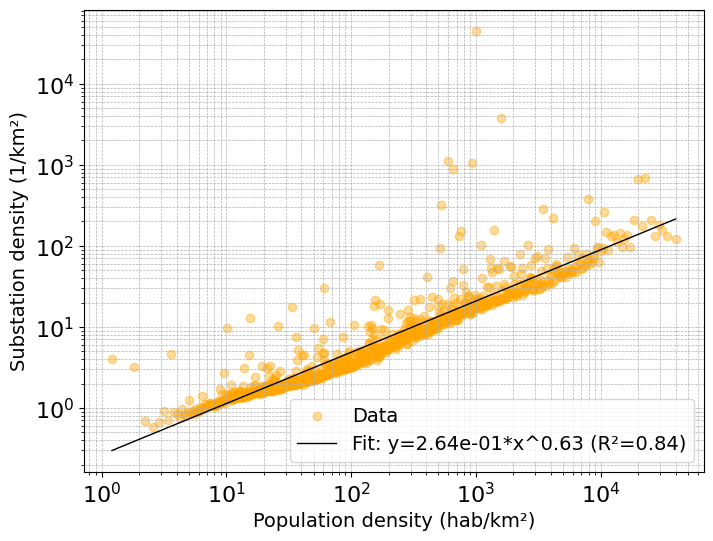}
    \caption{Same as \cref{fig:Nrho} but with a Voronoï cell layout.}
    \label{fig:voronoiNrho}
\end{figure}

We can optimize the cell layout using Voronoi cells as illustrated in \cref{fig:voronoi_mst}. Each Voronoi cell is centered on a substation $s$, and its boundaries delimit the fraction of the total area that is closer to the substation $s$ than to the other substations. Thus, each cell contains exactly one node, but varies in size. Given this cell layout, \cref{fig:voronoiNrho} illustrates that the strong rupture of the slope disappears. The two regimes appearing in \cref{regimes} seem to be artifacts of the artificial grid-cell lattice. The Voronoi grid seems to better capture the effective layout of substations. However, the fine-grained datasets that we use are based on a regular lattice spatial grid, and we cannot define the Voronoi tessellation without the location of nodes (which is our model output). This is why we have to keep the 2-regimes calibration.\\

\section{Correction to the minimum spanning tree}
\label{correction_appendix}

After we sample a spatial repartition of nodes, we estimate the kilometers of distribution lines needed to connect them. Since we are dealing with estimates only, we can use the \textit{minimum spanning tree} (MST) algorithm to connect the lines with the minimum total length. That said, the Kruskal process that we take is radical, and some corrections are needed to improve precision. \\

\begin{itemize}
    \item The MST lines are straight lines (distance "as the crow flies"). In reality, for practical reasons (like efficiency of building and maintenance), many distribution lines follow the road network \cite{arderne_predictive_2020} whose topology can be approximated by a square lattice . Assuming the distribution network maintains this square shape, we need to add a correction factor to account for the average rescaling between Manhattan and Euclidean distances. Let's consider an MST edge of Euclidean length $L_E$ at angle $\alpha$ relative to a square road grid (Figure 14). The Manhattan distance along the grid is 
\begin{equation}
L_M(\alpha) = (\cos \alpha +\sin \alpha ) L_E, 
\end{equation}
as depicted in \cref{fig:circle}. Averaging uniformly over $\alpha \in [0, \pi/2]$, this yields:
\begin{equation}
\langle L_M\rangle_\alpha = \frac{2}{\pi}\int_0^\frac{\pi}{2} L_M(\alpha ) \mathrm d \alpha=  \frac{4}{\pi} L_E.
\end{equation}
This leads to the first correction factor $\gamma_1 =\frac{4}{\pi}$. We emphasize that this correction is a rough approximation and not an exact correction. It assumes a uniform distribution of angles between the grid and the road and may fail on non-orthogonal road networks.\\

    \begin{figure}[h]
    \centering
    \includegraphics[width=0.7\linewidth]{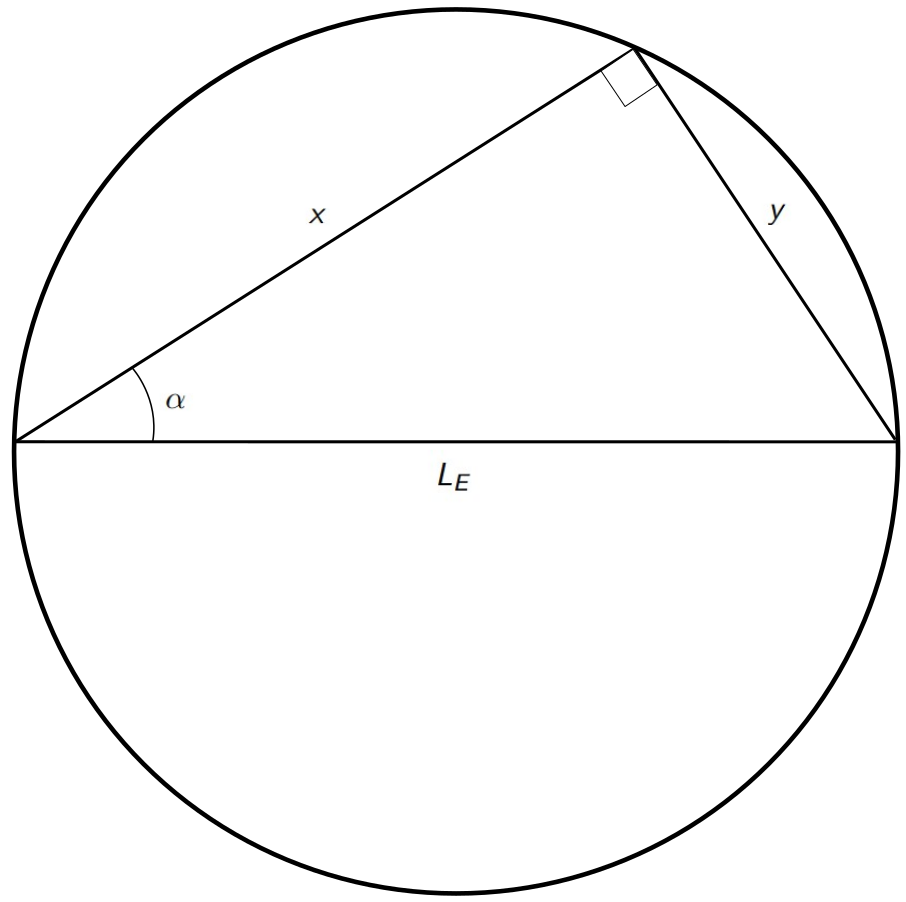}
    \caption{Drawing of the difference between Manhattan distance $L_M = x+y$ and Euclidean distance $L_E$ given an angle $\alpha$.}
    \label{fig:circle}
\end{figure}

    \item The uncorrected MST length value is for the routes (the pillar paths), which can contain more than one cable. To estimate material investments, we need to determine the cable length required for the simulated network. The ENEDIS open database provides access to line-shaped cable data as a set of areas corresponding to cable segments. For each cable segment of length $\ell_i$, the dataset contains a formal surface $a_i$ corresponding to the full surface within 5 meters of the line. Even though it is not rich enough to reconstruct a topology, this geographical data can be used to measure distances precisely. The total cable distance is the sum of all the line-shaped cable data:
\begin{equation}
L_\mathrm{cab} = \sum_{i=1}^{N_d} \ell_i,
\end{equation}
with $N_d$ the number of line-shaped entries. For the ENEDIS dataset, we get for the underground and aerial MV lines:
\begin{equation}
L_\mathrm{cab}^\mathrm{aer} = 330\;996 \;\mathrm{km}, \;\;\;\;\;\;L_\mathrm{cab}^\mathrm{und} = 376\;527 \;\mathrm{km}.
\end{equation}
This counts overlaps multiple times between the buffered areas. We take as a hypothesis the fact that when two cables share the same pillar path, then their buffered areas overlap. Using $|a|$ as the notation for the area of the surface $a$, we can estimate globally the total surface $A$ spanned by the $a_i$ with overlap included and the total surface $A'$ spanned by the $a_i$ with overlaps excluded (overlap will artificially increase the area spanned by cables): 
\begin{equation}
A = \sum_{i=1}^{N_d}  |a_i|, \;\;\;\;\;\;A' = \left| \bigcup_{i=1}^{N_d} a_i\right|.
\end{equation}
To factor out the overlap contribution in the total length estimates, we build a rescaled $L_\mathrm{route}$ such that 
\begin{equation}
L_\mathrm{route} = \frac{A'}{A} L_\mathrm{cab}.
\end{equation}
We now estimate the real-world length of the distribution with overlap excluded:
\begin{equation}
L_\mathrm{route}^\mathrm{aer} = 321\;629 \;\mathrm{km}, \;\;\;\;\;\;L_\mathrm{route}^\mathrm{und} = 299\;518 \;\mathrm{km}.
\end{equation}
We remark that most of the overlaps concern the underground edges: tunnels are dug to gather many cables. Having this done, we estimate the route-to-cable correction factor (calibrated on the French MV layer) as $\gamma_2 = \frac{A}{A'}\simeq 1.139$. The hypothesis chosen above relies on the reasonable scenario where buffer widths are small relative to inter-paths distances, but one that may break in urban areas where many tunnels can be superposed. We expect this assumption to slightly overestimate $\gamma_2$ in dense underground systems.\\

\item The MST is a lower bound; in reality, there are some bridge edges causing loops in the MV layer. To implement this partially looped structure, we can generalize the Kruskal algorithm as follows, taking as input a set of spatial nodes with their coordinates, and a parameter $b$. First, add all possible edges to form a clique network $G$ from the input node locations. Secondly, for each edge $(i, j)$ in decreasing length order, we check if removing $(i, j)$ will make the shortest path from $i$ to $j$ higher than $b$. If so, we keep $(i, j)$, otherwise, we delete it from the network. This has been done in a previous work \cite{emery_statistical_2025} to recover the missing length from the minimum spanning tree distribution grid model, where, comparing the data with the model, loops could explain the correction factor $\gamma_3 \simeq 1.060$. Even though the loops are an intuitive mismatch source, other effects can be the cause of the need for this residual correction factor.

\end{itemize}

All these informations could be included in a factor $\gamma = \gamma_1 \gamma_2 \gamma_3\simeq 1.537$ such that the total length of MV lines yields:

\begin{equation}
    L_\text{MV} = \gamma L_\text{MST}.
\end{equation}

An important note to consider is that $\gamma$ is partially calibrated on French ENEDIS data and is therefore not a universal correction. Applying it to other countries assumes that overlap structure ($\gamma_2$) and other effect like looping intensity ($\gamma_3$) are similar to the French case. These assumptions are supported by our cross-country results on $L_{\text{MV}}$ but not independently verified. Country-specific calibration would be preferable where local data permit. For the same reason, the French cable-length comparison in Table \ref{tab:FrenchIndicatorComp} should be read as a consistency check rather than a fully independent validation, since these correction factors enter its construction.

\section{Multiscale validation}
\label{multiscale}

We assess the spatial reliability of the model as a function of the aggregation scale by tiling metropolitan France into non-overlapping $k\times k$ blocks of grid cells ($k = 3, 5, 7, ...$) and computing, for each block containing at least one empirical and simulated data point, the mean squared logarithmic error between the two. This is the same metric used in the main text (Eq. \ref{MSElog}), now applied at the block level rather than the country level. We evaluate two quantities: the number of substations and the length of MV cables. Because the ENEDIS line dataset covers only the ENEDIS concession territory ($\simeq$~95\% of metropolitan France), the link-length comparison is restricted to blocks with a positive ground-truth ENEDIS length.

\begin{figure}
    \centering
    \includegraphics[width=1\linewidth]{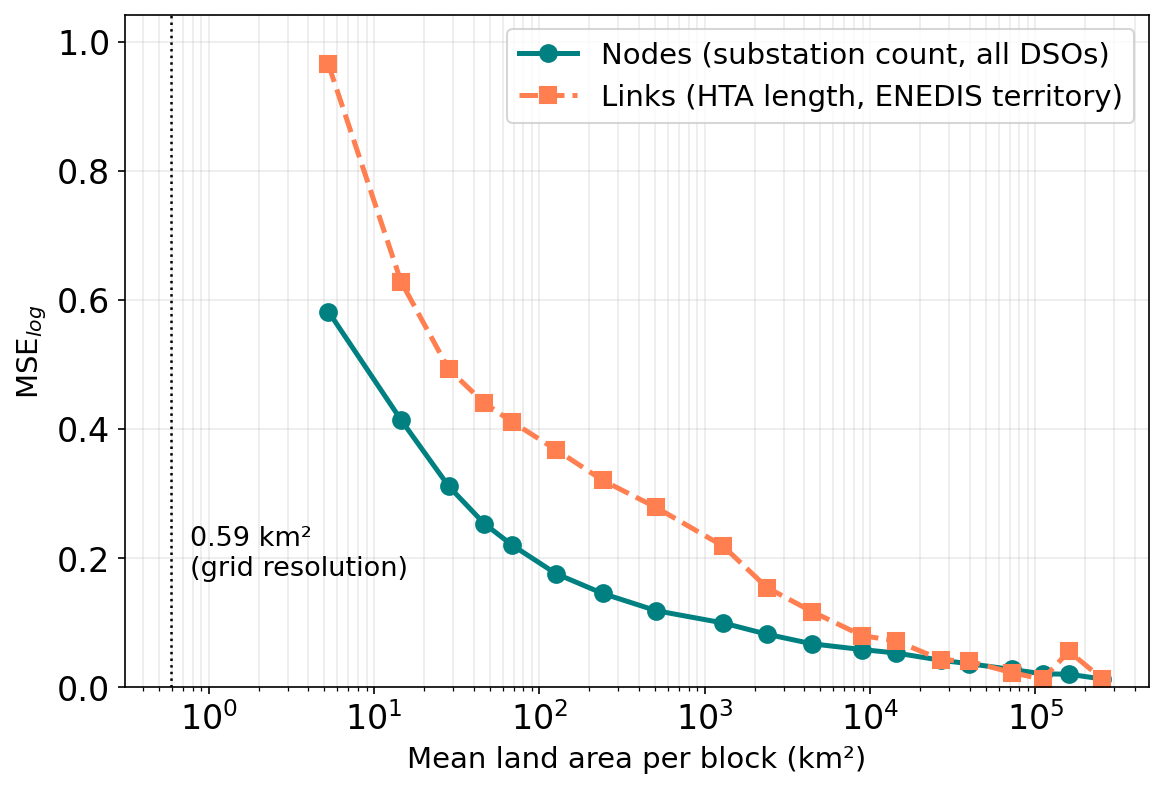}
    \caption{Multiscale validation of the model. The mean squared logarithmic error $\mathrm{MSE}_{\log}$ is shown as a function of the mean land area per aggregation block, for two model outputs: substation count and MV cable length. The block with $k = 1$ is omitted because only $\sim$28\% of cells are populated with nodes on both sides, thus leading to high bias.}
    \label{fig:multiscale}
\end{figure}

Figure \ref{fig:multiscale} shows $\mathrm{MSE}_{\log}$ as a function of mean land area per block for both substation count and MV cable length. Both curves decrease monotonically, demonstrating that local placement errors average out with aggregation scale. The convergence of both curves toward their respective global-bias values as $k$ increases demonstrates that the local discrepancies, dominated by stochastic effects, are canceled upon aggregation, as expected from a Poisson point process. Taken together, these results show that the model provides reliable spatial estimates above a few tens of km² and accurate national totals, while acknowledging that the 0.59 km² pixel remains a noisy floor resolution.

\section{Data sources}
\label{data_sources}

The sources of the literature values used in subsections \ref{subsectionB} and \ref{subsectionC} are detailed in table \ref{tab:grid_sources}.

\urldef{\urlCEER}\url{https://www.ceer.eu/wp-content/uploads/2024/04/7th-Benchmarking-Report-2022.pdf}
\urldef{\urlUKena}\url{https://assets.publishing.service.gov.uk/government/uploads/system/uploads/attachment_data/file/479267/clim-adrep-ena-2015.pdf}
\urldef{\urlOfgem}\url{https://www.ofgem.gov.uk/publications/2008-2009-electricity-distribution-quality-service-data-tables}
\urldef{\urlSlovakiaA}\url{https://www.tdworld.com/overhead-distribution/article/20962833/distributed-generation-drives-system-planning}
\urldef{\urlSlovakiaB}\url{https://www.ssd.sk/buxus/docs/dokumenty/o_nas/vyrocne_spravy/Vyrocna%20sprava%20a%20uctovna%20zavierka%202017%20SSD.pdf}
\urldef{\urlSlovakiaC}\url{https://www.atpjournal.sk/buxus/docs/dokumenty/kniznica_dokumentov/2014/Horak_ZSDS.pdf}
\urldef{\urlGermany}\url{https://www.geode-eu.org/wp-content/uploads/2024/06/Grids-for-Speed_Report.pdf}
\urldef{\urlEstonia}\url{https://www.energia.ee/-/doc/8457332/ettevottest/aastaaruanne2018/EE_Annual_report_2018.pdf}
\urldef{\urlUSAa}\url{https://www.energy.gov/sites/prod/files/2021/03/f83/Advanced%20Transmission%20Technologies%20Report%20-%20final%20as%20of%2012.3%20-%20FOR%20PUBLIC_0.pdf}
\urldef{\urlUSAb}\url{https://www.statplanenergy.com/wp-content/uploads/2021/08/Transformer-Report-Chapter-Summaries-TOC-Sample-Pages-Ed-9-2021.pdf}
\urldef{\urlUSAc}\url{https://www.energy.gov/sites/prod/files/2017/01/f34/Electricity%20Distribution%20System%20Baseline%20Report.pdf}
\urldef{\urlIndiaA}\url{https://www.ceicdata.com/en/india/electricity-number-of-transformers}
\urldef{\urlIndiaB}\url{https://now.solar/2021/04/07/growth-of-electricity-sector-in-india-from-1947-2020/}
\urldef{\urlChinaA}\url{https://ptr.inc/demystifying-chinas-power-grid/}
\urldef{\urlChinaB}\url{https://chinaenergyportal.org/en/2018-detailed-electricity-statistics-update-of-dec-2019/}
\urldef{\urlRussiaA}\url{https://www.cigre.org/userfiles/files/Community/National%20Power%20System/Pr%C3%A9sentation%20PowerPoint%20-%202020_National_power_system_RUSSIA.pdf}
\urldef{\urlRussiaB}\url{https://www.cigre.org/userfiles/files/Community/NC/Russia_The_Electric_Power_System.pdf}
\urldef{\urlMexicoA}\url{https://www.cenace.gob.mx/Docs/Planeacion/ProgramaRNT/Programa%20de%20Ampliaci%C3%B3n%20y%20Modernizaci%C3%B3n%20de%20la%20RNT%20y%20RGD%202019%20-%202033.pdf}
\urldef{\urlMexicoB}\url{https://www.proyectosmexico.gob.mx/wp-content/uploads/2018/09/VIIN-FINAL.pdf}
\urldef{\urlMexicoC}\url{https://base.energia.gob.mx/prodesen/PRODESEN2018/PRODESEN18.pdf}

\begin{table*}[b]
\centering
\small
\caption{Summary of data sources used for grid length and related indicators. All values originate from grey literature (regulatory, governmental, operator, and industry sources).}
\label{tab:grid_sources}
\setlength{\tabcolsep}{0pt}
\begin{tabular*}{\textwidth}{@{\extracolsep{\fill}}|l|l|l|}
\hline
\textbf{Country} & \textbf{Type of Data} & \textbf{Source / URL} \\ \hline

General & $L_{MV}$ & \parbox[t]{9cm}{\raggedright \urlCEER} \\ [2ex] \hline
General & \parbox[t]{4cm}{\raggedright Number of households, Electric energy consumption} & \parbox[t]{9cm}{\raggedright World Bank} \\ [2ex]\hline
Great Britain & $L_{MV}$, $N_{stat}$ & \parbox[t]{9cm}{\raggedright \urlUKena \par \urlOfgem} \\ [2ex]\hline
Slovakia &$L_{MV}$, $N_{stat}$ & \parbox[t]{9cm}{\raggedright \urlSlovakiaA \par \urlSlovakiaB \par \urlSlovakiaC} \\ [2ex] \hline
Germany & $N_{stat}$ & \parbox[t]{9cm}{\raggedright \urlGermany} \\ [2ex] \hline
Estonia & $N_{stat}$ & \parbox[t]{9cm}{\raggedright \urlEstonia} \\ [2ex]\hline
USA & $L_{MV}$, $L_{LV}$, $N_{stat}$ & \parbox[t]{9cm}{\raggedright \urlUSAa \par \urlUSAb \par \urlUSAc } \\ [2ex]\hline
India & $L_{MV}$, $L_{LV}$, $N_{stat}$ & \parbox[t]{9cm}{\raggedright \urlIndiaA \par \urlIndiaB} \\ [2ex]\hline
China & $L_{MV}$, $L_{LV}$, $N_{stat}$ & \parbox[t]{9cm}{\raggedright \urlChinaA \par \urlChinaB} \\ [2ex]\hline
Russia & $L_{MV}$, $L_{LV}$, $N_{stat}$ & \parbox[t]{9cm}{\raggedright \cite{kalt_global_2021}; \urlRussiaA \par \urlRussiaB } \\ [2ex]\hline
Mexico & $L_{MV}$, $L_{LV}$, $N_{stat}$ & \parbox[t]{9cm}{\raggedright \urlMexicoA \par \urlMexicoB \par \urlMexicoC} \\ \hline
\end{tabular*}
\end{table*}

\end{appendices}
\bibliography{biblio.bib}
\end{document}